\documentclass{article}
\usepackage{graphicx}
\begin{document}
\title{Analytical expression for wave scattering from exponential height correlated rough surfaces}
\author{
        M. Zamani$^{1}$, M. Salami$^2$, S. M. Fazeli$^3$, G. R. Jafari$^{1}$ \thanks{Email: g\_jafari@sbu.ac.ir} \\
        {\small $^1$ Department of Physics, Shahid Beheshti University, G.C., Evin, Tehran 19839, Iran} \\
        {\small $^2$ Department of Physics, Shahroud Branch, Islamic Azad University, Shahroud, Iran} \\
        {\small $^3$ Department of Basic Science, university of Qom, Qom, Iran}
        }

\date{\today}
\maketitle \baselineskip 24pt

\begin{abstract}
Wave scattering from rough surfaces in addition the inverse scattering
is an interesting approach to obtain the surface topography properties in various fields.
Analytical expression in wave scattering from some known rough
surfaces, not only help us to understand the scattering phenomena,
but also would prove adequate to be a criterion to measure the
information for empirical rough surfaces. For a rough surface with an
exponential height correlation function, we derive an analytical
expression for the diffused part and expanded it in two asymptotic
regimes. We consider one surface as slightly rough and the other as
very rough based on the framework of the Kirchhoff theory. In
the end, we have measured the role of various Hurst exponents and
correlation lengths on scattering intensity in self-affine surfaces.
We have shown that by increasing the Hurst exponent from $H=0$ to $H=1$,
the diffuse scattering decreases with the scattering angle.

\end{abstract}

\section{Introduction}

The study of rough surfaces is needed in many scientific fields.
This is due to the fact that roughness is a scaling parameter and
depending on the scale of observation which plays an important role
in natural phenomena. The behavior of scattered fields from rough
surfaces is of interest to may research topics, ranging from X-ray
to radar wave in a long time in various fields. One of the most
general approaches to study wave scattering is the Kirchhoff theory
\cite{beck,kong,Ogilvy,fung,Voronovich}. This theory is an
electromagnetic theory and is known as a 'tangent plane theory'
which is most widely used to calculate the distribution of the
specular and diffuse parts of the reflected light. The Kirchhoff
theory treats any point on a scattering surface as a part of an
infinite plane, parallel to the local surface tangent \cite{Ogilvy}.

There are two types of problems in this area: (a) direct problem,
(b) inverse problem. The direct problem is concerned with
determining the scattered field from the knowledge of the incident
field and the scattering obstacle
\cite{lu,simonsen,perez,caron,Shan,John}. Interesting works in the
context of wave scattering has been carried out Brewster's
scattering angle \cite{Brew} and the Rayleigh hypothesis \cite{Reig}
have been studied. In addition, Ingve et al \cite{Ingve} studied
wave scattering from self-affine surfaces where Leskova et al
\cite{Leskova} studied the coherence of p-polarized light scattered
from rough surfaces. Since the wave-length could be either smaller
or greater than the height fluctuations, the two scale theory was
proposed \cite{jafari}. In a further study, effects of interference
of two beam scattering was considered \cite{salami}. Scattering from
multilayer is studied by Carniglia \cite{Carniglia}. Elson et al
\cite{Elson} investigated vector wave scattering theory to estimate
the angular distribution of scattered light from optical surfaces.
Some of the researcher studied shadowing effect on rough surfaces
\cite{Fuks,Bruce}. The inverse problem is concerned with inverse
scattering techniques to \cite{liseno,ferraye,Simonsen2,Zhongxin}
measure the statistical properties of rough surfaces. Sinha et al
showed how scaling properties exists in X-rays and neutrons from
rough surfaces \cite{Sinhaprb}, Jafari et al \cite{jafari2} obtained
the surface roughness, Dashtar et al used inverse wave scattering to
determine the height distribution on a rough surface
\cite{dashtdar}.

The results indicate that when the wavelength is larger than the
correlation length, the scattered wave differs with the case where
the wavelength is smaller than the correlation length. In this work
an analytical expression for the diffused intensity is obtained. For
the case where the height correlation function has an exponential
behavior. In addition scattering from self-affine surfaces is
studied for various roughness exponents and correlation lengths.

The paper is organized as follows. In Sec. II we present a
theoretical description of Kirchhoff theory and surface roughness is
the source of scattered field. The self-affine fractal model of the
surfaces in Sec. III. The comparison of coherent and diffuse parts
of scattered intensity and scattering  from self-affine surfaces
with different parameters are described respectively in Sec. IV.
Finally, some general conclusions are presented in Sec. V.

\begin{figure}[t]
\centering
\includegraphics[width=8cm,height=7cm,angle=0]{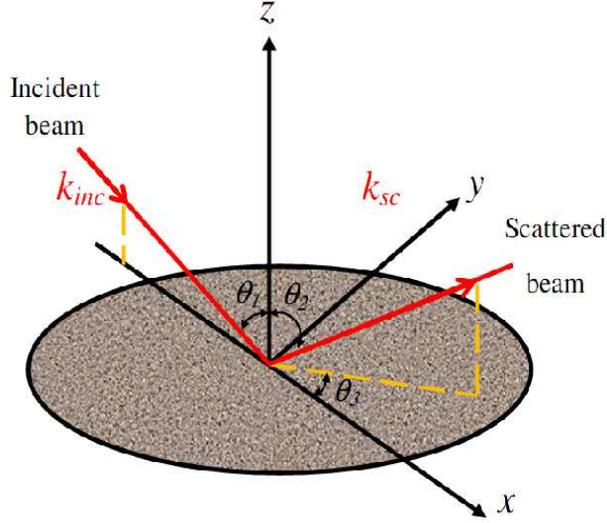}
\caption{Schematic figure to show the geometry used for wave scattering from a rough surface.} \label{Fig1}
\end{figure}

\section{Kirchhoff theory}

In Kirchhoff theory, the incident field is written as
$\psi^{inc}(r)=\mathrm{exp}(-i k_{inc}.r)$, where $k$ and $r$ are
the wave number and position respectively. Fig. 1 shows
schematically the scattering phenomena from a rough surface with
dimensions $-X\leq x_{0}\leq X$, $-Y\leq y_{0}\leq Y$. The scattered
field in $r$ is denoted by $\psi^{sc}$ and relies on three important assumptions: a) The surface is observed in the far
field. b) No point on the surface has infinite gradient. Therefore,
the Fresnel laws can be locally applied. c) The reflection
coefficient, $R_{0}$, is independent of the position on the rough
surface. The total scattered field over the mean reference plane
$A_{M}$, is given by \cite{Ogilvy},
\begin{eqnarray}
\psi^{sc}(r) &=& \frac{ike^{ikr}}{4\pi r}\int_{S_{M}}
(a\frac{\partial h}{\partial x_{0}}+b\frac{\partial h}{\partial
y_{0}}-c)
\times \exp(ik[Ax_{0}+By_{0}+ Ch(x_{0},y_{0})]) dx_{0} dy_{0}, \\
A&=&\sin \theta_{1}-\sin \theta_{2} \cos \theta_{3}, \nonumber \\
B&=&-\sin \theta_{2} \sin \theta_{3}, \nonumber \\
C&=&-(\cos \theta_{1} +\cos \theta_{2}), \nonumber \\
a&=&\sin \theta_{1}(1-R_{0})+\sin \theta_{2}\cos
\theta_{3}(1+R_{0}), \nonumber \\
b&=&\sin \theta_{2} \sin \theta_{3}(1+R_{0}), \nonumber \\
c&=&\cos \theta_{2}(1+R_{0})-\cos \theta_{1}(1-R_{0}). \nonumber
\end{eqnarray}

The coherent and the diffuse intensities are given by \cite{Ogilvy},
\begin{eqnarray}
I_{coh} &=& I_{0}e^{-g}, \nonumber \\
<I_{d}> &=& \frac{k^{2}F^{2}}{2\pi r^{2}}A_{M} e^{-g} \int
_{0}^{\infty}J_{0}(kR\sqrt{A^{2}+B^{2}}) \times[e^{gCor(R)}-1] R dR,
\end{eqnarray}
where $I_0$ is the scattered intensity from a flat interface. The
parameter $g$ is equal to $=k^{2}\sigma^{2} C^{2}$, where $C=\cos
\theta_{1} + \cos \theta_{2}$, and the height-height correlation
function $Cor(R)$ which is definite for an isotropic surface
$(\langle h(x) \rangle=0)$ is equal to $\frac{\langle h(x+R)
h(x)\rangle}{\sigma^2}$.

\section{Scattering from self-affine surfaces}
\subsection{Self-affine surface}

The correlation function of self-affine surfaces is known by
correlation length, $\xi$, which is the mean lateral length of
surface features, e.g. grain size or other.Following Sinha et al.
\cite{Sinhaprb,Palasantzas,Sahimi}, can be approximated by an
analytic function, $Cor(R)\approx e^{-(\frac{R}{\xi})^{2H}}$, which
contains the following limits. For $R\gg \xi$, the correlation
vanishes, $(Cor(R)=0)$ and for $R \ll \xi$, the correlation function
would look like $Cor(R)\simeq 1-(\frac{R}{\xi})^{2H}$, and roughness
exponent $H$ would be between zero and one. The large values of $H$
correspond to smoother height-height fluctuations, while small
values of $H$ characterize irregularity in height for rough surfaces
at the length scales ($R\ll \xi$)
\cite{Sinhaprb,Palasantzas,Sahimi}.

For a Gaussian correlation function ($H=1$), Eq. (2) yielded an
expression for the diffuse scattered intensity \cite{Ogilvy}:
\begin{eqnarray}
<I_{d}>=\frac{k^{2}F^{2}\xi^{2} e^{-g}}{4\pi r^{2}}A_{M}
\sum_{n=1}^{\infty}\frac{g^{n}}{n! n} \exp(-\frac{k^{2}(A^{2}+
B^{2})\xi^{2}}{4n}).
\end{eqnarray}

\begin{figure}[t]
\centering
\includegraphics[width=12cm,height=9cm,angle=0]{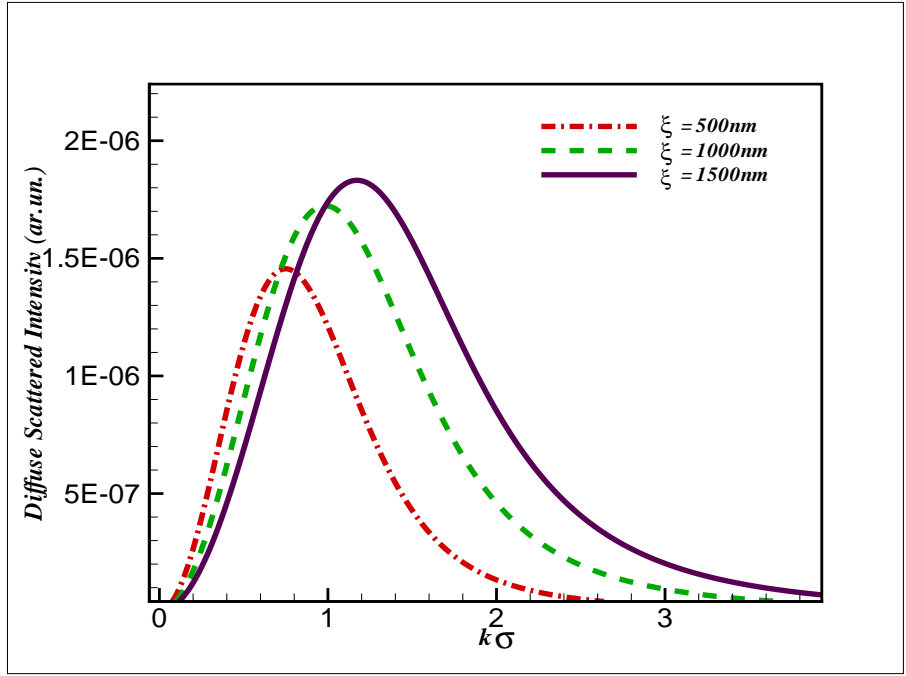}
\caption{(Color online) Dependence of diffuse scattered intensity on
$k\sigma$ for different correlation length ($\xi=500$, $1000$,
$1500nm$) and for angles $\theta_{1}=\theta_{3}=0^{\circ}$,
$\theta_{2}=20^{\circ}$, wavelength $\lambda=500 nm$ and $H=0.5$.}
\label{fig2}
\end{figure}
\begin{figure}[t]
\centering
\includegraphics[width=12cm,height=14cm,angle=0]{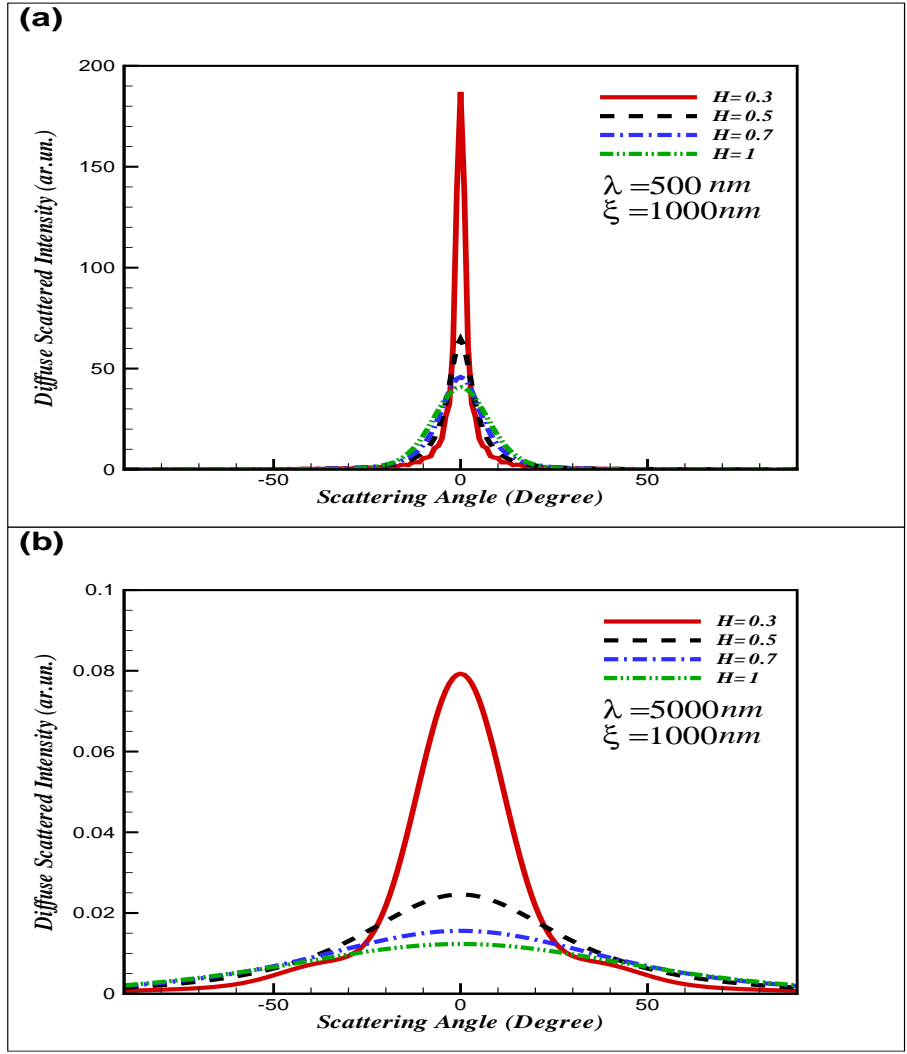}
\caption{(Color online) Dependence of diffuse scattered intensity on
$\theta_{2}$ for angles $\theta_{1}=0^{\circ}, \theta_{3}=0^{\circ}$
and different Hurst exponent, correlation length $\xi=1000 nm$,
standard deviation $\sigma=50 nm$ and the incident wavelength (a)
$\lambda=500 nm$, (b) $\lambda=5000nm$.} \label{fig3}
\end{figure}
\begin{figure}[t]
\centering
\includegraphics[width=12cm,height=9cm,angle=0]{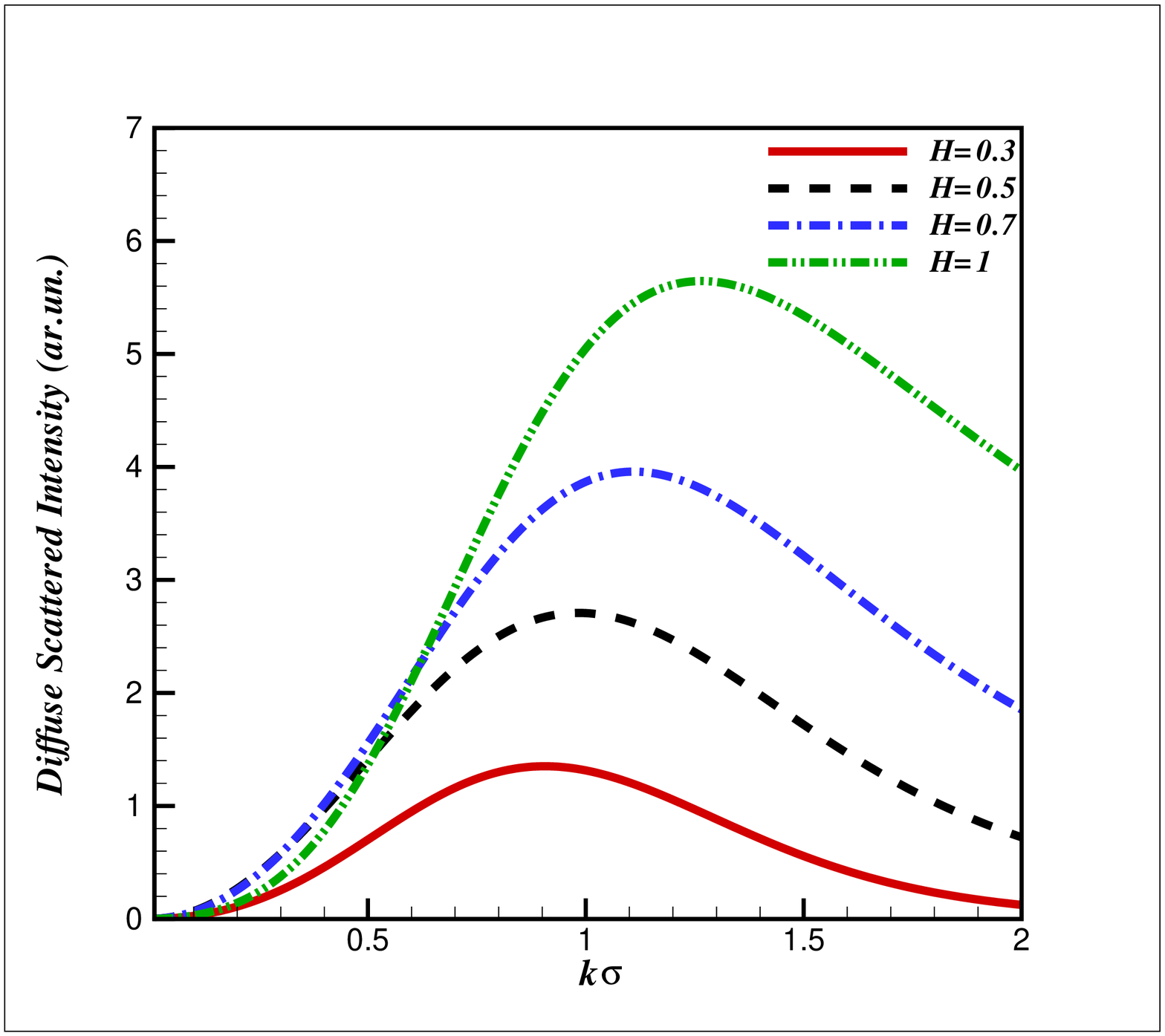}
\caption{(Color online) Dependence of diffuse scattered intensity on
$k\sigma$ for angles $\theta_{1}=\theta_{3}=0^{\circ}$ and
$\theta_{2}=20^{\circ}$ different Hurst exponent, correlation length
$\xi=1000 nm$, wavelength $\lambda=500 nm$.} \label{fig4}
\end{figure}

\subsection{Analytical expression for a rough surface with exponential height correlated}

Knowing any exact results in wave scattering from rough surfaces
is helpful to better understand the scattering phenomena. One of these
surfaces is exponential height correlated surface which could be an
important and suitable criteria to measure
information content in other rough surfaces. We obtain an analytical
expression for the correlation function $Cor(R)=
e^{-(\frac{R}{\xi})^{2H}}$ with $H=0.5$, which appears in an
exponential form $Cor(R)=e^{-\frac{R}{\xi}}$. With regarding $ e^{g
Cor(R)}=\sum_{n=0}^{\infty}\frac{g^{n}cor^{n}(R)}{n!},$ the diffuse
part of the scattered intensity will be:
\begin{eqnarray} <I_{d}>&=&\frac{k^{2}F^{2}}{2\pi r^{2}}A_{M} e^{-g}
\int _{0}^{\infty}J_{0}(kR\sqrt{A^{2}+B^{2}})
\times\sum_{n=1}^{\infty}\frac{g^{n}cor^{n}(R)}{n!} R dR \nonumber\\
&=& \frac{k^{2}F^{2}}{2\pi r^{2}}A_{M} e^{-g}
\sum_{n=1}^{\infty}\frac{g^{n}}{n!} \int
_{0}^{\infty}J_{0}(kR\sqrt{A^{2}+B^{2}})cor^{n}(R) R dR \nonumber\\
&=& \frac{k^{2}F^{2}}{2\pi r^{2}}A_{M} e^{-g}
\sum_{n=1}^{\infty}\frac{g^{n}}{n!} \int
_{0}^{\infty}J_{0}(kR\sqrt{A^{2}+B^{2}}) e^{-\frac{nR}{\xi}} R dR
\nonumber\\&=&\frac{k^{2}F^{2}}{2\pi r^{2}}A_{M} e^{-g}
\sum_{n=1}^{\infty}\frac{g^{n}}{n!} \frac{\xi^{2}}{n^{2}
(1+\frac{k^{2}(A^{2}+B^{2})\xi^{^{2}}}{n^{2}})^{\frac{3}{2}}}.
\end{eqnarray}
Eq. (4) enables the study of rough surfaces in two asymptotic limits:

\textbf{A. Slightly rough surfaces}, For slightly rough surfaces $(g \ll 1)$,
the series in Eq.(4) converges quickly and only the first term needs to be
considered. So, the diffuse field intercity becomes:
\begin{equation} <I_{d}>=\frac{k^{2}F^{2}}{2\pi r^{2}}A_{M} e^{-g} g
\frac{\xi^{2}}{
(1+k^{2}(A^{2}+B^{2})\xi^2)^{\frac{3}{2}}}.
\end{equation}
\textbf{B. Very rough surfaces}, The total intensity is $\langle
I\rangle=\langle I_{coh}\rangle+\langle I_d\rangle=e^{-g}I_0+\langle
I_d\rangle$, when the surface is very rough ($g\gg1$), the coherent
field will be negligible and the total intensity is equivalent to
the diffuse ones \cite{Ogilvy} (Chap. 4, Page 92). In addition, we
know that $I_{0} \propto X.\frac{Sin(kAX)}{kAX}$, when sample size
is larger than wavelength, the coherent field appears just in
specular angle. For the non-specular angle we have only diffuse
field ($\lambda\ll X,Y$ or $kAX\gg 1 \Rightarrow
X.\frac{Sin(kAX)}{kAX}=\pi.\delta(kA)$) \cite{Glazov}. So, according
Eq.(2) the total scattered intensity which is approximately equal to
one could be written as:
\begin{eqnarray}
<I>=\frac{k^{2}F^{2}}{2\pi r^{2}}A_{M}
\int_{0}^{\infty}J_{0}(kR\sqrt{A^{2}+B^{2}}) e^{-g} (e^{gCor(R)}-1)
R dR.
\end{eqnarray}

When the length scale is larger than the correlation length
($R\gg\xi$), the correlation function vanishes ($Cor(R\gg\xi)=0$).
In this case, according to Eq. (6), there is no scattered intensity
in the large length scale ($R\gg\xi$). For convergence, it is
sufficient to extend the numerical integration from zero to  $5\xi$.
In the category which $g \gg 1$ and for small range of $R$, Eq. (6)
is equal to:
\begin{eqnarray}
<I>&=&\frac{k^{2}F^{2}}{2\pi r^{2}}A_{M} \int
_{0}^{\infty}J_{0}(kR\sqrt{A^{2}+B^{2}}) e^{-g [1-Cor(R)]}R dR
\nonumber\\ &=& \frac{k^{2}F^{2}}{2\pi r^{2}}A_{M} \int
_{0}^{\infty}J_{0}(kR\sqrt{A^{2}+B^{2}}) e^{-g\frac{R}{\xi}} R dR \nonumber\\
&=&\frac{k^{2}F^{2}}{2\pi
r^{2}}A_{M}\frac{\xi^{2}}{(1+\frac{k^{2}(A^{2}+B^{2})\xi^{2}}{g^{2}})^{\frac{3}{2}}g^{2}}.
\end{eqnarray}


\section{Effects of the correlation length and Hurst exponent on scattering intensity}

\subsection{Variation of Correlation Length}

According to Eq. (2) the correlation function only effects the diffuse
part leaving the coherent part uneffected. Fig. 2 compares
the diffused scattering vs $k \sigma$, with a fixed $\lambda= 500$
 nm, for three values of the correlation length $\xi=500,1000,1500$
 nm and the angles $(\theta_{1}=\theta_{3}=0^{\circ},
\theta_{2}=20^{\circ})$. We have chosen the value of parameters in
the way which are compatible with experimental setup. The numerical
results which were obtained from Eq. (2) are shown for $k \sigma=0$
to $4$. A characteristic feature seen in Fig. 2 is the presence of a
maximum in the diffused part as a function of the $k \sigma$, which
depends on correlation length. By increasing the correlation length
the peak of the diffused part is shifted to large $k\sigma$. Indeed,
small fluctuation have low effects on larger wavelengths. Note that
decreasing wavelength and increasing rms have the same effect on
wave scattering. Figuratively speaking, a surface with shorter
correlation length appears rough to small wavelengths and smooth to
large wavelengths. Therefore, in the context of scattering, $k
\sigma$ is more suitable than $k$ or $\sigma$ alone. In other words
the longer angles would play a more significant role on diffused
intensity. This result in a decrease$\backslash$increase of
intensity for smaller$\backslash$longer angles. In addition, the
smaller correlation length changes its behavior as it shifts to
larger $k\sigma$.

\subsection{Variation of Hurst exponent}

The Hurst exponent appears in the correlation function and the
correlation function enters into the diffuse part. So, the
variations of Hurst exponent could effect the diffuse scattered
intensity, too. To more sense the role of Hurst exponent in wave
scattering, we study diffuse scattering in two regimes. Fig. 3
depicts the variation of the diffuse scattering intensity with
respect to scattering angle for different Hurst exponents for the
two cases: $\lambda < \xi$ \& $\lambda > \xi$. The first point in
Fig. 3 is that by increasing $H$, the diffuse scattering intensity
is reduced for angles close to the specular angle widening the
curve. Thus it can be observed in larger angles. For non-specular
scattering angle far from specular angle, this behavior is opposite,
so by increasing $H$, the diffuse scattering intensity is increased.
Another point is that for $\lambda < \xi$ the fluctuations of the
surface are observed better and the roughness is increased so the
diffuse scattering intensity increases (Fig. 3a). For the case
$\lambda > \xi$, smaller fluctuations of the surface are not
observed and surface seems smoother and the diffuse scattering
intensity decreases sharply. This behavior can be seen by comparing
Figs. 3a and 3b. Variation of the diffuse scattering intensity with
respect to $H$ has the same behavior for both wavelengths smaller
and larger than the correlation length.

There are two kinds of behaviors in scattering intensity based on
wavelength smaller or larger than the correlation length. When the
wavelength is less than correlation length, the small height
fluctuation is being observed. So the wavelength acts as a scale for
the observation of the surface. The more observed height fluctuation
leads to the more diffuse scattered intensity. There is opposite
behavior when the wavelength of incident wave is larger than the
surface correlation length. In this case, the incident wavelength is
not able to recognize small fluctuations on the surface and loses
the scale of scattering angle. This means that the surface seams
smooth for this scale of observation (wavelength). This results in a
decrease in the diffuse intensity.

Fig. 4 shows the variation of the diffuse scattering intensity with
respect to $k\sigma$ for different $H$ for the non-specular
scattering angles ($\theta_{1}=\theta_{3}=0^{\circ}$ and
$\theta_{2}=20^{\circ}$). By increasing $H$, the diffuse scattering
intensity decreases, which is in good agreement with Fig. 3.
In Fig. 4, by increasing $H$, the curve shifts to larger $k\sigma$.

In order to investigate the role of the Hurst exponent, a surface
with the observed values of for the correlation length ($\xi=1000
nm$), and wavelength ($\lambda=500 nm$) is considered. When $k
\sigma$ is small, either by means of small $\sigma$ or large
$\lambda$, the scattering is not sensitive to the corrugation,
independently of the Hurst exponent. For the chosen correlation
length of $1000$ nm, the diffuse intensity in Fig. 4 is similar for
$k \sigma < 0.4$, the influence of the Hurst exponent becomes
apparent only for large $k \sigma$.


\section{Conclusion}

The roughness of a surface effects the scattered wave from rough
surfaces. We have found an analytical expression for spectral scattered
intensity for a rough surface with exponential height correlation
function and investigated the results for asymptotic regimes of
slightly rough and very rough surfaces. The exact solution helps us
understand the scattering phenomena and could be a suitable
criteria to measure the value of information in a unknown rough
surface.

The role of the Hurst exponent has been investigated for self-affine
rough surfaces. Such an examination was performed over a wide range of surface
topographies, from logarithmic ($H =0$) to a power-law self-affine
rough surface, $0<H<1$. The roughness exponent $H$ has a strong
impact on the diffused part of wave scattering mainly for relatively
large correlation lengths. Therefore, Hurst exponent must be taken
carefully into account before deducing the roughness correlation lengths
from wave scattering measurements.

{\bf Acknowledgements:} The authors would like to thank the research
council of Islamic Azad University of Shahroud for financial support
and Soheil Vasheghani Farahani for helping to edit the manuscript.

\end{document}